\documentclass[11pt]{article}
\usepackage[utf8]{inputenc}
\usepackage{consilr2024}
\usepackage[table]{xcolor}
\usepackage{makecell}
\usepackage{tabularx}
\usepackage{array} 
\usepackage{times}
\usepackage{url}
\usepackage{latexsym}
\usepackage{authblk}
\usepackage{hyperref}
\usepackage{etoolbox}
\usepackage{graphicx}
\usepackage{subcaption}
\usepackage{subcaption}
\usepackage{float}
\usepackage{url}

\newcommand{\keyword}[1]{
\hspace{0.2cm}%
\fontsize{10}{12}\selectfont%
\textbf{Keywords: } %
}

\title{\textbf{BRIDGING NEUROSCIENCE AND AI: ADAPTIVE, CULTURALLY SENSITIVE TECHNOLOGIES TRANSFORMING APHASIA REHABILITATION}}
\author[1] {Andreea I. Niculescu}
\author[2]{Jochen Ehnes}
\author[3]{Minghui Dong}

\affil [1]{A*STAR Institute for Infocomm Research (I2R)
\break
\texttt{andreeaniculescu@sigchi.org}}

\break

\affil[2]{A*STAR Institute for Infocomm Research (I2R)
 \break
\texttt{jehnes@gmail.com}}

\affil[3]{A*STAR Institute for Infocomm Research (I2R)
 \break
\texttt{dongminghui@gmail.com}}

\date{}
\begin{document}
\maketitle
\begin{abstract}
Aphasia, a language impairment primarily resulting from stroke or brain injury, profoundly disrupts communication and everyday functioning. Despite advances in speech therapy, barriers such as limited therapist availability and the scarcity of personalized, culturally relevant tools continue to hinder optimal rehabilitation outcomes. This paper reviews recent developments in neurocognitive research and language technologies that contribute to the diagnosis and therapy of aphasia. Drawing on findings from our ethnographic field study, we introduce two digital therapy prototypes designed to reflect local linguistic diversity and enhance patient engagement. We also show how insights from neuroscience and the local context guided the design of these tools to better meet patient and therapist needs. Our work highlights the potential of adaptive, AI-enhanced assistive technologies to complement conventional therapy and broaden access to therapy. We conclude by outlining future research directions for advancing personalized and scalable aphasia rehabilitation.
\end{abstract}

\section{Introduction}
\label{intro}
Aphasia is a complex language disorder, typically caused by stroke or brain injury, that significantly impairs an individual’s ability to communicate. It affects core language functions—speaking, understanding, reading, and writing—and has profound implications for quality of life. Each year, approximately 15 million people suffer strokes, and up to 40\% develop aphasia \cite{WHO}.

This condition can seriously hinder communication, making it difficult for individuals to express themselves or understand others, often leading to social isolation and emotional distress \cite{hilari2009health} \cite{parr2007living}. Aphasia presents in various forms, each affecting language differently. Broca’s aphasia (non-fluent) involves slow, effortful speech with preserved comprehension, while Wernicke’s aphasia (fluent) results in fluent but often nonsensical speech and impaired understanding \cite{damasio1992aphasia}. Another major subtype is \textit{Primary Progressive Aphasia} (PPA), a neurodegenerative condition marked by gradual atrophy in language-related brain regions. Unlike stroke-induced aphasia, PPA stems from progressive diseases like frontotemporal dementia or Alzheimer’s, and it worsens over time. It often begins with subtle word-finding issues and progresses to more severe impairments, while other cognitive functions remain initially intact. PPA includes three variants—nonfluent/agrammatic, semantic, and logopenic—each with distinct linguistic and neural profiles  \cite{mesulam2003primary}.

Currently, no medical or surgical treatments exist for aphasia, making speech
therapy the primary intervention. Research has shown that consistent speech therapy can lead to significant improvements in patients’ conditions \cite{roberts2021better}. 
Rehabilitation programs are tailored to address the specific challenges experienced by individuals with different types of aphasia. These highly personalized interventions target each patient’s unique communication needs, often strengthened through a collaborative therapeutic alliance between the clinician and the individual. To guide therapy effectively, clinicians use assessments such as the \textbf{Boston Diagnostic Aphasia Examination (BDAE)} to evaluate language abilities and design customized treatment plans \cite{marie2024validity} \cite{cherney2011intensity}. The success of aphasia therapy depends on factors such as aphasia type, severity, patient motivation, cognitive profile, and therapy intensity, with evidence indicating that early, intensive, and tailored interventions yield the best outcomes \cite{berthier2005poststroke} \cite{biel2022motivation}.

Despite its importance, many individuals receive insufficient therapy. Barriers such as therapist shortages, limited insurance coverage, mobility constraints, and caregiver availability, along with traditional sessions—often lasting up to two hours daily for optimal results—pose further challenges \cite{stahl2018efficacy}, \cite{guo2014speech}. Tele-rehabilitation emerged as a promising way for patients to get therapy from home, helping to overcome distance or other practical problems \cite{towey2012telepractice}. Computer programs can also help with rehabilitation, training skills like speaking, reading aloud, writing, understanding, memory, and finding words \cite{katz2010computers} \cite{loverso1992microcomputer}. However, their repetitive nature and lack of personalization often reduce patient motivation over time \cite{niculescu2023assistive}.

This paper integrates insights from neuroscience, language technology, and clinical practice, presenting findings from our fieldwork and prototype development of culturally and linguistically adaptive tools for aphasia rehabilitation. Specifically, we explore the integration of neurolinguistic principles and natural language processing (NLP) methods into assistive technology to support speech therapy. Our work focuses on developing a prototype tailored to the needs of aphasia patients in Singapore, aiming to bridge the gap between clinical intervention and home-based practice through adaptive, user-centered design.
%
%
    %
    %
    %
    %
    %
    %

\section{Background and related work}
Breakthroughs in neuroscience have significantly deepened our understanding of the neural foundations of language, while innovations in language technologies are opening new avenues for diagnosis and intervention, positioning aphasia research at the intersection of multiple disciplines. This section provides a two-part review: the first part surveys cognitive neuroscience research that underpins current therapeutic approaches; the second examines emerging developments in language technology—from automatic speech recognition and AI-driven assessments to digital therapy platforms—that are reshaping diagnostic and treatment practices.

\subsection{Neuroscience insights}
Progress in neuroscience and neurolinguistics reveals how the brain processes language, the ways aphasia disrupts these functions, and possible pathways for recovery. Neuroplasticity plays a key role in how the brain reorganizes after injury, supported by insights from neuroimaging and electrophysiological markers that track language deficits and recovery. Additionally, recognizing that language involves multifunctional brain networks highlights the importance of addressing cognitive impairments alongside linguistic rehabilitation.

\subsubsection{Neuroplasticity and non-invasive brain stimulation (NIBS)}
A main focus in neuroscience and language research is \textbf{neuroplasticity}, the brain’s remarkable ability to reorganize and form new connections after injury. Brain imaging shows that the brain can compensate for language problems by recruiting alternative paths, often in the right hemisphere or areas near the injury. Understanding these compensatory processes helps researchers learn how the brain adapts after stroke or neurodegenerative diseases \cite{turkeltaub2025right}. Based on this knowledge, researchers are developing personalized rehabilitation programs tailored to each person’s type of aphasia and brain injury. Studies show that therapy matched to a patient’s neural profile is more effective at promoting recovery \cite{kiran2019neuroplasticity}.

Building on this, \textbf{non-invasive brain stimulation (NIBS)}—including transcranial direct current stimulation (tDCS) and repetitive transcranial magnetic stimulation (rTMS)—is being used alongside speech therapy. These techniques modulate brain activity and support the reorganization of language networks \cite{Fridriksson2018} \cite{sebastian2016tDCS}. Clinical trials suggest that tDCS, combined with therapy, could help patients with both post-stroke and primary progressive aphasia (PPA) improve skills such as naming, fluency, and communication \cite{Meinzer2016}\cite{Tsapkini2014}. Research on post-stroke aphasia also suggests tDCS may reduce mental fatigue and help patients maintain attention, supporting longer-term benefits \cite{Fridriksson2018} \cite{Kang}.

\subsubsection{Neuroimaging and electrophysiological markers}
Advanced neuroimaging techniques, including functional MRI (fMRI), diffusion tensor imaging (DTI), and magnetoencephalography (MEG), are crucial for understanding how the brain reorganizes after stroke, particularly in chronic aphasia \cite{stockert2020dynamics} \cite{li2022fmri}. These methods are often combined with electrophysiological techniques such as event-related potentials (ERPs), which provide detailed timing information about language processing. 

Two ERP components—N400 and P600—are especially informative. The N400, a negative deflection around 400 milliseconds after a word is presented, reflects \textbf{semantic processing}, or the retrieval and integration of word meaning. It occurs relatively quickly because understanding meaning relies on direct access to memory and context. The P600, a positive deflection around 600 milliseconds, is linked to \textbf{syntactic processing}, or the analysis of grammatical structure, which is computationally more complex and therefore slower. In aphasia, the N400 is often delayed or reduced, indicating semantic impairments, while the P600 may be diminished or absent, signaling syntactic deficits. Importantly, both components can change with therapy: successful treatment can increase N400 amplitude and normalize its timing, while re-emergence of the P600 may indicate recovery of syntactic integration \cite{stockert2022therapy}. These ERP patterns can therefore be used to diagnose language impairments and to track progress during rehabilitation, serving as sensitive neural markers of both deficits and neuroplasticity in aphasia recovery.

\subsubsection{Neural multifunctionality and cognitive impairments}
While these neurophysiological markers provide important insights into core language processes, research increasingly shows that aphasia often affects broader cognitive functions beyond language. Studies in neurolinguistics have found that deficits in attention, theory of mind, and executive function frequently occur alongside morphosyntactic, semantic, and pragmatic difficulties \cite{Caplan2020}. These  cognitive impairments are associated with the severity of aphasia and with structural changes in the brain, particularly in frontotemporal regions \cite{StrikwerdaBrown2019} \cite{Fittipaldi2019}. For example, individuals with primary progressive aphasia (PPA) may show early problems with emotion recognition and empathy, even when language abilities remain relatively intact. Similarly, post-stroke aphasia is linked to reduced performance on attention and executive function tasks, which closely relates to communication difficulties \cite{Murray2012}. These findings support the concept of neural multifunctionality—the idea that language operates within broader domain-general brain systems—and highlight the importance of interventions that target both linguistic and cognitive aspects of aphasia \cite{cahana2014} \cite{Kytnarova}.

From this broader perspective, pharmacological interventions are being investigated as complements to behavioral therapy, with the potential to improve not only language deficits but also co-occurring cognitive impairments \cite{stockbridge2022better}. Although still in early clinical stages, some compounds aim to enhance cerebral blood flow, modulate neurochemical balance, or support neural recovery, thereby promoting improvements in both language and cognitive function \cite{Berthier03062021} \cite{Mayo}. As understanding of the neural basis of language and cognition grows, pharmacological strategies may become an increasingly important component of personalized, integrated aphasia rehabilitation programs.

\subsection{Language technology developments}
Language technologies are increasingly important for supporting aphasia diagnosis, therapy, and communication. Advances in artificial intelligence (AI) have made these tools more scalable, adaptive, and personalized, complementing traditional clinical care. They can be used for automated assessment and diagnosis, personalized or adaptive therapy, and augmentative or alternative communication (AAC) to help patients communicate when natural speech is severely limited.
\subsubsection{Apahsia assessment and diagnosis}
AI technologies are increasingly used for the early detection, accurate diagnosis, and ongoing monitoring of aphasia. Machine learning models can now extract detailed linguistic and acoustic features from speech to identify language impairments with high precision. For example, \cite{Zusag2023} developed an automated speech recognition (ASR) and natural language processing (NLP) pipeline that classified aphasia subtypes with up to 90\% accuracy—comparable to expert clinicians. Similarly, \cite{Herath2022} used deep neural networks and Mel Frequency Cepstral Coefficients (MFCCs) to classify ten levels of aphasia severity with over 99\% accuracy, showing promise for more precise severity grading and personalized therapy planning.

Beyond classification, AI models are increasingly being applied to fine-grained error analysis and discourse-level interpretation. \cite{Salem2023} fine-tuned a large language model (LLM) on picture-elicited narratives to identify target words in paraphasic speech, achieving 50.70\% accuracy. While further work is needed, this study demonstrates how AI can automatically detect and analyze paraphasic expressions in connected speech, taking a step closer to automatic aphasic discourse analysis.

AI is also advancing longitudinal tracking in progressive language disorders such as PPA. Subtle changes in lexical diversity, syntactic complexity, or speech rate can signal cognitive decline. As noted by \cite{Zhong2024}, AI-integrated systems offer non-invasive, cost-effective tools for continuous monitoring, supporting earlier intervention and personalized therapy adjustments.

\subsubsection{Augmentative \& alternative communication}

Language technologies are increasingly providing new ways for people with aphasia to communicate, through both AI-enhanced Augmentative and Alternative Communication (AAC) systems and emerging Brain–Computer Interfaces (BCIs). Modern AAC devices now use machine learning and natural language processing to generate contextually appropriate phrases and predictive text. For example, Transformer-based models can adapt vocabulary to a user’s communication history and conversational context, improving fluency and reducing cognitive effort \cite{su2025review}. User-centered studies also show that features such as real-time grammar support and visual feedback make communication more accurate and boost user confidence \cite{mao2025hic}. 

At the same time, non-invasive BCIs are showing promise in decoding imagined speech—translating neural activity into words. Recent EEG-based research highlights major progress in signal processing, feature extraction, and classification accuracy, especially using attention-based deep learning models \cite{su2025peerj}, \cite{panachakel2021decoding}. Some clinical trials have begun using BCI-driven neurofeedback in aphasia therapy. For instance, \cite{Musso} found that EEG-based feedback during word detection tasks improved language abilities across several domains in chronic post-stroke aphasia, suggesting these systems may both restore communication and enhance neural plasticity.

Together, these developments point toward a shared goal: bridging the gap between thought and expression for individuals with severe speech or motor impairments. Although most BCI applications are still experimental, the combination of adaptive AI and brain-based communication technologies holds great promise for improving expressive communication in aphasia related disorders.

\subsubsection{Language technologies for aphasia therapy}
Recent advances in language technologies—such as automatic speech recognition (ASR), neural machine translation (NMT), and natural language processing (NLP)—are increasingly being applied to aphasia therapy to strengthen pronunciation, grammar, vocabulary, and overall communication skills. For example, in a small study of an iPad-based speech therapy app, the ASR system correctly identified whether participants pronounced each word accurately in the same way as human raters about 80\% of the time, and participants showed improved word production after four weeks of ASR-guided practice \cite{ballard2019}. NMT models have also been used to automatically translate aphasic sentences into fluent ones (BLEU $\approx$ 38), demonstrating the feasibility of providing grammatical and lexical support to aid communication \cite{smaili2022language}.

In addition, integrated ASR–NLP approaches can transcribe spontaneous aphasic speech with a word error rate of 24.5\% and, when combined with feature-based classifiers, can detect aphasia with 86\% accuracy \cite{barberis2024automatic}.

Building on these capabilities, mobile and gamified apps have made therapy more accessible and engaging for long-term self-directed use. Virtual reality app, such as \cite{bu2022mobile}, offer immersive language and cognitive training targeting oral expression, comprehension, and cogntive skills; they have been pilot-tested with patients, healthy volunteers, and clinicians. Commercial programs including \cite{vasttx2017}, \cite{constanttherapy2023}, \cite{bungalow2021}, \cite{aphasiascripts2021}, \cite{sentencesharper2023}, and \cite{ucl2024italkbetter} offer structured exercises targeting multiple language and cognitive skills (reading, writing, speaking, and problem-solving). Many now use adaptive algorithms that adjust difficulty based on performance \cite{Zhong2024}.

Last but not least, Melodic Intonation Therapy (MIT) remains an important complementary intervention for non-fluent aphasia. By using melody, rhythm, and pitch to stimulate speech, it has shown effectiveness in both subacute and chronic stages. A randomized controlled trial conducted across multiple therapy centers reported significant gains in repetition and communication after six weeks of MIT \cite{vandermeulen2014efficacy}. A crossover trial also confirmed functional improvements using the Communicative Activity Log (CAL), a standardized tool measuring how well patients communicate in daily life rather than only in the clinic\cite{haro2019pilot}. Neuroimaging evidence shows that MIT promotes right-hemisphere reorganization and increases structural connectivity of the \textit{arcuate fasciculus} in patients with non-fluent aphasia \cite{zhang2023melodic}. New approaches are now combining MIT with ASR-based prosody feedback, though these innovations have yet to be fully evaluated.
\section{An aphasia training tool for Singaporean patients}
\subsection{Motivation}
Despite advances in language technologies and the growing availability of digital tools for home-based aphasia therapy, significant limitations remain—particularly in generalizing rehearsed speech to spontaneous, real-world communication \cite{simmons2014}. Many existing systems still lack adaptive feedback, contextual relevance, cultural and linguistic personalization. Motivated by these challenges, our team—drawing on expertise in speech recognition under noisy and reverberant conditions \cite{Chong}, \cite{Robot_Olivia}, \cite{Robot_Framework}, \cite{Smart_City}, \cite{IDA_Interact}, \cite{TechForFuture}, as well as in chatbot development across various domains \cite{niculescu2020digimo}, \cite{niculescu2009stress} — engaged in a focused collaboration with Tan Tock Seng Hospital in Singapore to develop assistive tools to local patient needs. 
  
\subsection{Field study} 
To gain a nuanced understanding of the challenges faced by both patients and clinicians in aphasia therapy, we conducted an ethnographic field study. The study included structured observations of therapy practices, qualitative analysis of data samples from the Open AphasiaBank corpus \cite{aphasiabank2017}, and interviews with speech-language pathologists from Tan Tock Seng Hospital (TTSH). The results of this study, published in \cite{niculescu2023assistive} identified several structural and interactional challenges in aphasia therapy. Patients struggled with sentence construction, word retrieval, and pronunciation. Therapists, in turn, emphasized the importance of individualized therapy plans incorporating storytelling, conversational practice, grammatical exercises, and music-based tasks—such as singing familiar phrases—to activate automatic speech sequences. 

Motivating patients, adapting materials to local cultural contexts, and supporting long-term independent training were recurrent themes. Therapists frequently used mental connections between words and images to stimulate speech production and relied on family input to establish a core vocabulary drawn from the patient’s environment and everyday routines. However, technological limitations—such as ASR tools that fail with non-American accents or therapy apps that use culturally irrelevant imagery—continue to hinder effective therapy delivery.

Further interviews highlighted that local patients present wide variation in language proficiency, literacy levels, and dominant languages (e.g., English, Malay, Chinese dialects, Tamil, etc.). Therapy must often be adapted to suit these differences, especially when literacy impairments or apraxia of speech are present. Yet current therapy apps frequently rely on Western-centric content or accent-dependent recognition, limiting their usability in the local context.

As patients improve, they seek more advanced training—such as more complex sentence construction—but lack appropriate digital tools beyond therapist-guided sessions. The absence of structured feedback during home practice poses a risk of reinforcing errors, especially for individuals living alone. These insights underscore the need for culturally relevant, feedback-rich tools that can support both patients and therapists in tracking progress and promoting recovery.

\subsection{Mapping neurolinguistic and contextual insights to therapy goals and prototype features}

Drawing on ethnographic findings and neurocognitive research on aphasia, we mapped key neurolinguistic and contextual insights to corresponding NLP applications and speech therapy goals, highlighting potential solution strategies realized through prototype implementation. This mapping guided the design of two digital therapy prototypes, emphasizing targeted, adaptive, and linguistically grounded interventions. Table~\ref{tab:prototype-alignment} summarizes these connections, with implemented features highlighted in \textbf{bold} and planned features in \textit{italic}, showing how advances in speech recognition and NLP are applied to support personalized and engaging home-based rehabilitation that complements traditional clinical therapy.

\setlength{\extrarowheight}{3pt} 
\begin{table}[h!]
\centering
\footnotesize
\setlength{\tabcolsep}{4pt} 
\renewcommand{\arraystretch}{1.1} 

\rowcolors{2}{gray!10}{white}
\begin{tabularx}{\textwidth}{|>{\raggedright\arraybackslash}p{3cm} 
                          |>{\raggedright\arraybackslash}p{3cm} 
                          |>{\raggedright\arraybackslash}p{3cm} 
                          |>{\raggedright\arraybackslash}X|}
\hline
\makecell{\rule{0pt}{14pt} \textbf{Neurolinguistic} \\ \textbf{insights}} & 
\makecell{\textbf{NLP} \\ \textbf{applications}} & 
\makecell{\textbf{Speech therapy goals}} & 
\makecell{\textbf{Prototype implementation}} \\
\hline

\makecell[l]{\rule{0pt}{14pt}1. Knowledge of \\ syntactic deficits} & 
Use parsers to detect malformed sentences & 
\makecell[c]{\rule{0pt}{14pt}Sentence-level \\ training} & 
\textbf{Prototype 2:} 'Complete the Sentence’ targets syntax via sentence construction (1:1 match) \\

\hline

\makecell[l]{\rule{0pt}{14pt}2. Lexical-semantic \\ mapping in the brain} & 
\makecell[c]{Semantic vector \\ models} & 
\makecell[c] {Suggest related words \\during naming} & 
\textit{Planned:} Future versions of both prototypes may use LLMs to offer flexible suggestions \\

\hline

\makecell[l]{\rule{0pt}{14pt}3. Multimodal associa-\\tion in lexical access} & 
\makecell[c]{Vision-language \\ models} & 
\makecell[c]{Use image cues to \\ trigger speech} & 
\textbf{\rule{0pt}{14pt}Prototype 1:} Current 1:1 image - word pronunciation match; visual stimuli planned \\

\hline

\makecell[l]{\rule{0pt}{14pt}4. Real-time language \\ processing deficits} & 
\makecell[c] {ERP-based dialogue \\pacing}& 
Adapt turn-taking pace & 
\textit{Planned:} Potential for future interaction design \\

\hline
5. Pronunciation errors & \makecell[c]{\rule{0pt}{14pt} Classification models \\ on aphasic speech} & \makecell[c]{\rule{0pt}{14pt}Error-specific \\ feedback} & 
\textbf{Prototype 1:} Custom scoring + DTW detect phonetic errors, give personalized feedback \\

\hline

\makecell[l]{\rule{0pt}{14pt}6. Reduced speech \\ fluency \& word retrieval} & 
\makecell[c]{ \rule{0pt}{14pt}Rhythmic/formulaic \\language models} & 
\makecell[c]{ \rule{0pt}{14pt}Support automaticity;\\ use fixed patterns} & 
\textit{Planned:} Singing-based or patterned repetition exercises may be added \\
\hline
\makecell[l]{ \rule{0pt}{14pt}7. Storytelling \& \\ conversation deficits} & 
\makecell[c] {\rule{0pt}{14pt} Use LLMs} & 
\makecell[c] { \rule{0pt}{14pt}Support long \\ sentence formation}  & 
\textit{Planned}: GPT model embeddings \\
\hline

\makecell { \rule {0pt}{14pt}\textbf{Other contextual} \\ \textbf{ insights} } & 
\makecell [c]{\rule{0pt}{14pt}\textbf{Strategies} }  & 
\makecell [c]{\rule{0pt}{14pt}\textbf{ Speech therapy goals}} & 
\makecell[c]{\vspace{2pt} \rule{0pt}{14pt}\textbf{Prototype implementation}} \\
\hline

\hline

\makecell[l]{\rule{0pt}{14pt}1. Lack of personal \\ \& cultural adaptation} & 
\makecell[c]{\rule{0pt}{14pt} Enable caregiver\\ involvement} & 
\makecell[c]{\rule{0pt}{14pt} Engage patients \\ with relevant content} & 
\textbf{Prototype 2:} Exercises tailored to personal, cultural context via caregiver input \\

\hline

\makecell[l]{\rule{0pt}{14pt}2. Lack of remote\\ progress monitoring} & 
\makecell[c]{\rule{0pt}{14pt} Enable therapist \\ involvement } & 
\makecell[c] {\rule{0pt}{14pt} Track improvement \\ outside the clinic} & 
\textbf{Prototype 1:}  pronunciation scoring \textbf{ Prototype 2:} Session logs + therapist view \\

\hline

  \makecell[l]{\rule{0pt}{14pt}3. Fatigue  \&  repetitive\\ exercise routine} 
& \makecell[c]{\rule{0pt}{14pt} Gamification}
& \makecell[c]{\rule{0pt}{14pt} Increase patient \\ engagement} & 
\textbf{Prototype 2:} Uses tree-growth metaphor to encourage consistent training\\
\hline

\makecell[l]{\rule{0pt}{14pt}4. Local accent \\ challenges in ASR} & 
\makecell[c]{\rule{0pt}{14pt} Accent-adapted \\ ASR models} & 
\makecell[c]{\rule{0pt}{14pt}Broader accessibility \\ for diverse speakers} & 
\makecell[l]{\textbf{Prototype 1:} uses ASR adapted \\ to Singaporean English} \\

\hline
\end{tabularx}

\caption{Mapping neurolinguistic and contextual insights to speech therapy goals and prototype features}
\label{tab:prototype-alignment}
\end{table}

\subsubsection{Prototype 1}
\begin{figure}[htbp]
  \centering
  \includegraphics[width=1.0\linewidth]{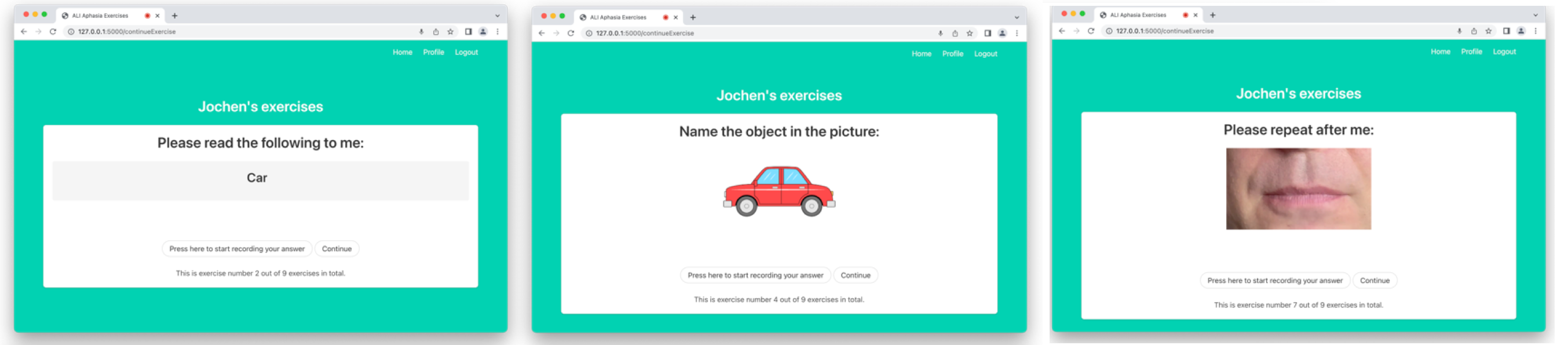}  
  \caption{Prototype 1 – exercises for word reading, object naming, and pronunciation with lip sync}
  \label{fig:aphasia-prototype}
\end{figure}

The first prototype, as presented in \cite{niculescu2023assistive}, aims to facilitate home-based speech training for individuals with mild aphasia by addressing the lack of immediate and personalized feedback in existing digital interventions. It is a web-based system that employs multimodal prompting—comprising text, images, audio, and synchronized lip videos — to accommodate patients with varying degrees of aphasia and to enhance comprehension of therapy tasks (see Figure \ref{fig:aphasia-prototype}).

Each word is paired one-to-one with an image cue. Patients are shown an image and invited to speak the name of the object. They can practice reading, naming, listing, and repetition. At the end of each task, patients receive a \textbf{pronunciation score} along with an \textbf{explanation of errors}, and can replay a lip-synchronization video to review the \textbf{correct pronunciation}.\footnote{A video demonstration of the prototype is available here: https://www.youtube.com/watch?v=JkL7lXQm7z8}.

At the core of the system is a speech recognition engine adapted specifically for Singaporean-accented English to better reflect the phonetic characteristics of the target user population. While British English is often preferred in applications for its perceived authority and prestige \cite{VoiceAccents}, Singapore English—a local variety influenced by Chinese and Malay—is commonly used in everyday interaction.

The initial speech-to-text conversion facilitates basic content matching, while a secondary evaluation layer assesses pronunciation accuracy using Dynamic Time Warping (DTW) \cite{senin2008dtw}. This method compares patient utterances against multiple prerecorded templates, enabling detailed and nuanced feedback that accommodates accent variability and low-resource language conditions.

To enhance semantic generalization—particularly in object-naming tasks—the system compares recognized speech against a set of acceptable alternative responses, minimizing false-negative evaluations. While the current semantic matching operates via rule-based logic, future iterations aim to incorporate large language models (LLMs) to support more natural and flexible user input.

The system is cross-platform, supports patient progress tracking, and include \textbf{therapist involvement}.

\subsection{Prototype 2}
Building on the foundational Prototype 1, Prototype 2, \textbf{SpeakNow}, was designed and implemented by \cite{BingHong2024} as a web-based speech therapy platform featuring dual interfaces: one for patients engaging in therapy, and another for caregivers and therapists to create, assign, and monitor exercises.

\begin{figure}[htbp]
  \centering
  \begin{subfigure}[b]{0.3\linewidth}
    \centering
    \includegraphics[width=\linewidth]{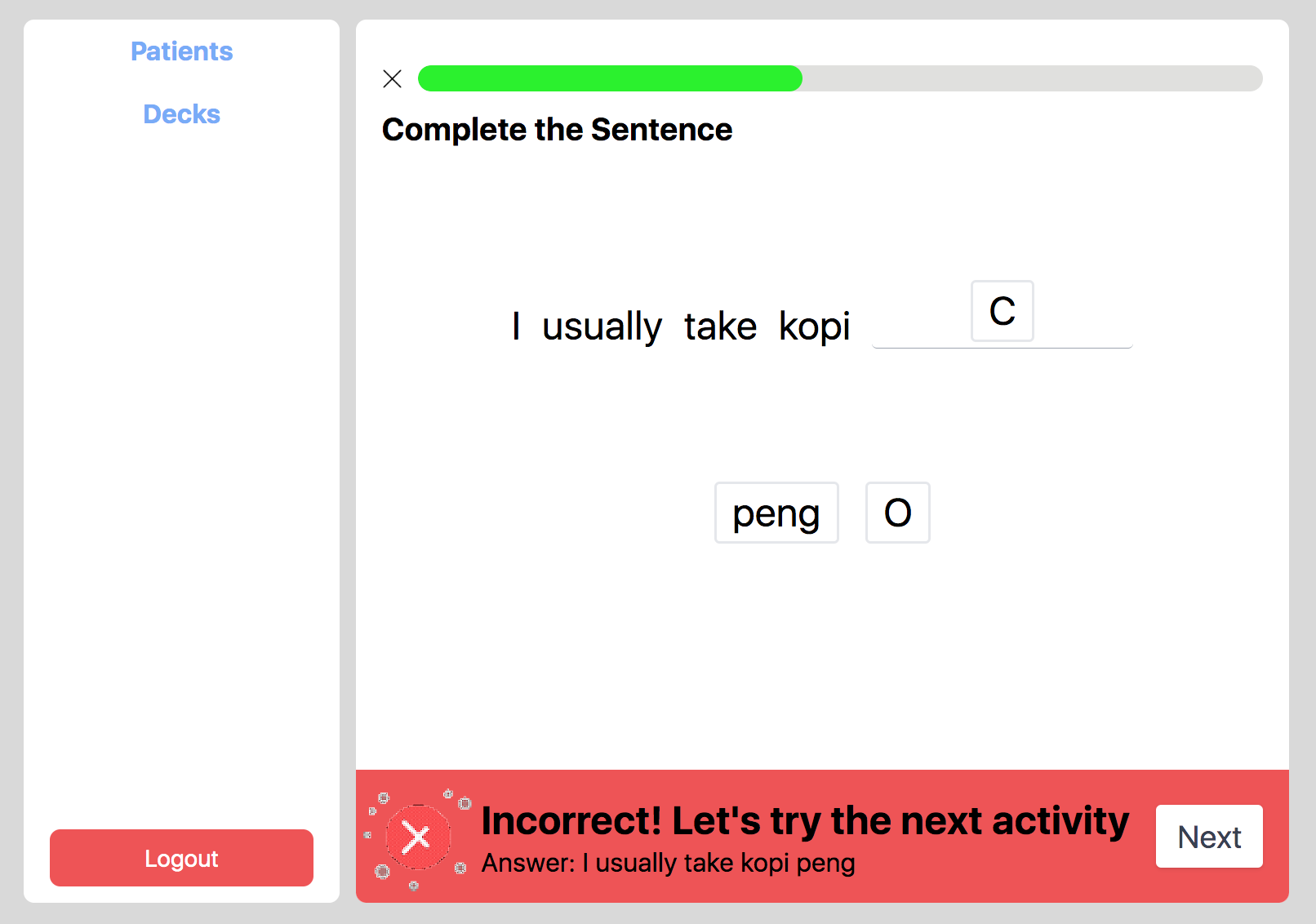}
    \caption{Complete Sentence}
    \label{Complete_Sentence}
  \end{subfigure}
  \hfill
  \begin{subfigure}[b]{0.3\linewidth}
    \centering
    \includegraphics[width=\linewidth]{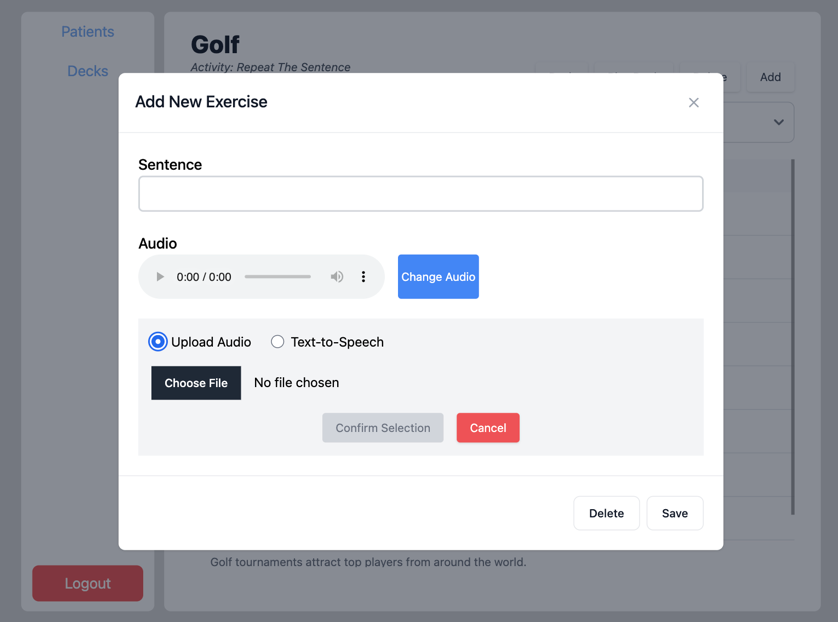}
    \caption{Adding a Repeat Sentence item}
    \label{Repeat_Sentence}
  \end{subfigure}
  \hfill
  \begin{subfigure}[b]{0.3\linewidth}
    \centering
    \includegraphics[width=\linewidth]{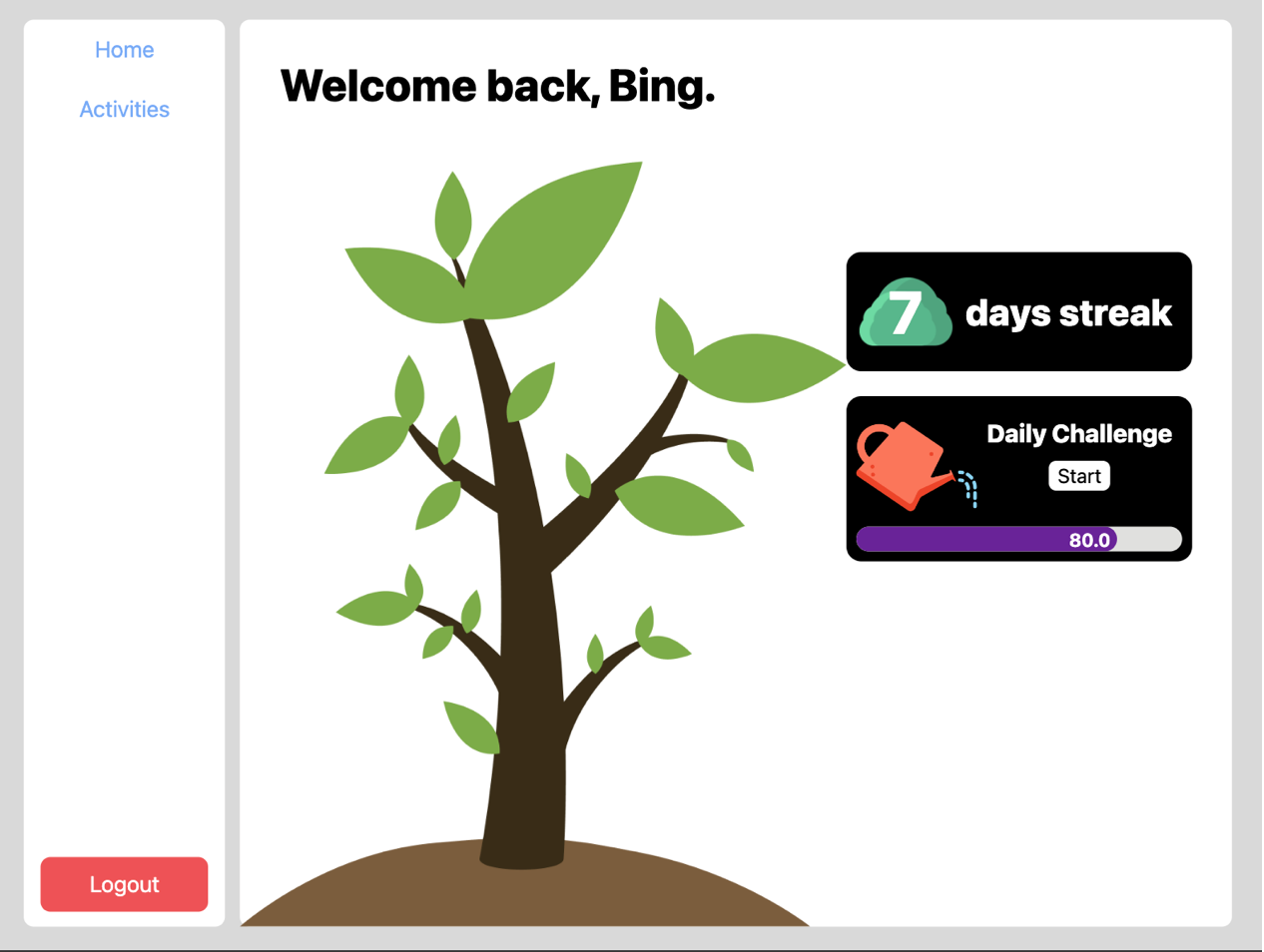}
    \caption{Virtual Tree}
    \label{Virtual_tree}
  \end{subfigure}
  \caption{Prototype 2 screenshots}
  \label{fig:three-images}
\end{figure}

A key feature is the customizability of therapy decks, which allows caregivers to design exercises tailored to the patient’s \textbf{personal history}, \textbf{cultural background}, and \textbf{language proficiency}. For instance, practice materials can reference familiar local dishes, hawker foods, or recognizable streets and neighborhoods in Singapore, alongside meaningful life events. For example, as shown in Figure \ref{fig:three-images}a, one exercise refers to Kopi C—a popular local coffee made with evaporated milk and sugar—reflecting everyday cultural references familiar to many Singaporeans. This personalization enhances emotional relevance and motivation by grounding therapy in the patient’s lived experiences. 

Therapy exercises are organized into decks, which can be created manually or generated with ChatGPT-assisted input, with audio integrated via text-to-speech (TTS). Each deck may include multiple exercise types, such as \textbf{Complete the Sentence} (Figure \ref{Complete_Sentence}) and \textbf{Repeat the Sentence} (Figure \ref{Repeat_Sentence}).

\textbf{Complete the Sentence} targets syntax and sentence construction through 1:1 word-to-blank matching. Patients reconstruct incomplete sentences by dragging and dropping words into blanks, with distractors included to increase task complexity. In the future, we plan to allow users to construct their own sentences based on given scenarios. To support this, we are exploring the integration of automatic parsers to detect lexical and syntactic errors, extending beyond simple rule-based matching.

\textbf{Repeat the Sentence} focuses on improving pronunciation and auditory comprehension. Patients listen to a sentence—either generated using Google Text-to-Speech or manually uploaded—and repeat it aloud. Real-time automatic speech recognition (ASR) analyzes the spoken response and provides feedback based on a predefined similarity threshold (set at 75\%).

To encourage daily engagement, caregivers can assign personalized tasks, which patients access via a \textit{Daily Challenge} feature. The system selects exercises randomly from pre-set decks. All user interactions—including accuracy, duration, and frequency—are logged in MongoDB and visualized through the therapist dashboard. Performance summaries are generated at the end of each session and can be reviewed daily or weekly. This feature supports \textbf{remote monitoring} and \textbf{adaptive planning}.
The interface is minimalistic and aphasia-friendly, with large, clearly labeled elements accommodating motor and cognitive impairments \cite{janeback2020designing}. Therapists can register patients, link caregiver accounts, and manage therapy routines.

To promote consistent engagement, \textbf{SpeakNow} employs lightweight gamification grounded in the Octalysis framework \cite{Gellner2022}. A virtual tree metaphor represents progress: completing daily tasks nurtures its growth, while missed sessions reset it, tapping into the emotional drive to avoid loss (Figure \ref{Virtual_tree}). Instant feedback via encouraging messages further reinforces engagement.

Built with the MERN stack and hosted on AWS, \textbf{SpeakNow} supports media APIs, secure storage, and scalable backend infrastructure.

\section{Future Directions and Conclusions}
Our work represents an early but important step toward AI-enhanced assistive technologies for aphasia therapy in Singapore. A key limitation is that, although our prototypes have been conceptually and technically validated, they have not yet been tested with patients. Meaningful progress in aphasia rehabilitation depends on iterative, user-centered evaluations, which we plan to conduct once approvals are granted. Without this feedback, no technology—however sophisticated—can fully meet users' complex needs.

Our ethnographic findings have provided valuable insights into everyday therapy challenges, helping us shape tools that reflect local linguistic diversity, cultural context, and patient motivation. These insights guided the design of our second prototype, \textbf{SpeakNow}, which supports functional communication through personalized, adaptive, and culturally grounded therapy experiences.

Looking ahead, we plan system-level enhancements, including replacing rule-based parsing with syntactic parsers to detect malformed utterances; integrating semantic vector and visual-to-language models to suggest related words during naming tasks; and incorporating music-based therapy elements such as rhythmic repetition and singing to stimulate formulaic language patterns. We also intend to experiment with LLM-based conversational agents to train pragmatic and turn-taking skills—bridging structured exercises and spontaneous conversation. Once implemented, these enhancements will be evaluated with patients for usability, engagement, and therapeutic effectiveness.

Advancing aphasia research will require diverse and representative datasets, particularly for non-English and disfluent speech, to improve the robustness of automatic speech recognition and language modeling. Future systems should embrace multilingual and multimodal interaction—including gestures, facial expressions, and prosody—to better mirror real-world communication. Integrating neurocognitive markers, such as ERP-based recovery indicators, into adaptive therapy algorithms is another promising direction requiring close collaboration between AI researchers and neuroscientists.

Despite rapid progress, most current technologies lack real-time adaptability, personalized feedback, and cultural sensitivity. Few systems have been evaluated for long-term engagement or sustainability in clinical and home settings. Addressing these gaps requires interdisciplinary collaboration and continuous co-design with patients and clinicians. Future systems that combine clinical insight with adaptive, feedback-driven designs can more effectively help individuals with aphasia regain functional communication.
\bibliographystyle{consilr}
\bibliography{consilr2024}

\end{document}